\documentclass[aps,prb,twocolumn,a4paper,showpacs]{revtex4-1}

\usepackage{graphicx}
\usepackage{dcolumn}
\usepackage{bm}
\usepackage{amsmath}
\usepackage{MnSymbol}%
\usepackage{color}

\begin{document}

\preprint{APS/123-QED}

\title{Impact of Oxygen Diffusion on Superconductivity in YBa$_2$Cu$_3$O$_{7-\delta}$ Thin Films Studied by Positron Annihilation Spectroscopy}

\author{M. Reiner$^{1}$, T. Gigl$^{1,3}$, R. Jany$^2$, G. Hammerl$^2$,  and C. Hugenschmidt$^{1,3,*}$}

\affiliation{$^1$Lehrstuhl E21 at Physics Department, Technische Universit\"at M\"unchen, James-Franck Stra\ss e 1, 85748 Garching, Germany}
\affiliation{$^2$Experimental Physics VI, Center for Electronic Correlations and Magnetism, University of Augsburg, Universit\"atsstra\ss e 1, 86135 Augsburg, Germany}
\affiliation{$^3$ FRM\,II at Heinz Maier-Leibnitz Zentrum (MLZ), Technische Universit\"at M\"unchen, Lichtenbergstra\ss e 1, 85748 Garching, Germany}

\date{\today}
 
\begin{abstract}
The oxygen deficiency $\delta$  in YBa$_2$Cu$_3$O$_{7-\delta}$ (YBCO) plays a crucial role for affecting high-temperature superconductivity.
We applied (coincident) Doppler broadening spectroscopy of the electron-positron annihilation line to study in situ the temperature dependence of the oxygen concentration and its depth profile in single crystalline YBCO film grown on SrTiO$_3$ (STO) substrates.
The oxygen diffusion during tempering was found to lead to a distinct depth dependence  of $\delta$, which is not accessible using X-ray diffraction.
A steady-state reached within a few minutes is defined by both, the oxygen exchange at the surface and at the interface to the STO substrate.
Moreover, we revealed  the depth dependent critical temperature $T_{\mathrm{c}}$ in the as prepared and tempered YBCO film. \\

\itshape{\footnotesize{$^*$Corresponding author: christoph.hugenschmidt$@$frm2.tum.de}}
\end{abstract}


\maketitle

\section{Introduction}
The cuprate YBa$_{\text{2}}$Cu$_{\text{3}}$O$_{\text{7}-\delta}$ (YBCO) is one of the most prominent representatives of oxides exhibiting High-Temperature Superconductivity (HTS).\cite{PhysRevLett.58.908,1347-4065-26-4A-L314,bednorz} 
The structure of YBCO is characterized by  two different lattice sites specific for the Cu atoms which are the chain site Cu(1) and the plane site Cu(2).
Oxygen deficiency in YBCO, i.e. the presence of oxygen vacancies at the chain site for $\delta >0$, strongly affects the hole doping within the CuO$_2$ planes. 
This hole doping, however, drives exceptional phenomena highly relevant for technical applications and fundamental solid state physics such as HTS with a critical temperature $T_\mathrm{c}$ above 90\,K and, presumably intertwined, Charge-Density-Wave (CDW) order .\cite{11827562,Ghiringhelli17082012,changnature} 
For this reason, in literature the oxygen deficiency $\delta$ is widely used for the characterization of the superconducting properties of YBCO crystals.

For technical applications such as superconducting wires, fault current limiters or magnets as well as for the fundamental investigation of the interplay between CDW and HTS phases high-quality crystals of YBCO are required. 
Since oxygen diffusion in YBCO is strongly affected by temperature heat treatment in vacuum or oxygen atmospheres has become a standard procedure for adjusting the value of $\delta$. 
Hence, the applied preparation process finally settles the distribution of oxygen in the YBCO sample.
However, even in single-crystalline thin films a clear impact of the tempering procedure on the lateral homogeneity of $\delta$ was observed.\cite{ybcoapl} 

In this paper we report on the depth dependent investigation of the oxygen distribution in YBCO films and the evolution of $\delta$  during tempering. 
We applied (Coincident) Doppler Broadening Spectroscopy ((C)DBS) of the electron-positron annihilation line using a slow positron beam\cite{RevModPhys.60.701, beam} in order to determine the depth profile of the oxygen deficiency. 
After implantation and thermalization, positrons diffuse through the crystal until they annihilate with electrons at a typical rate of around $5\cdot10^{11}$s$^{-1}$ in singe-crystalline YBCO.\cite{RevModPhys.66.841,0953-8984-1-23-020} 
In this system positrons show a particular high affinity to the oxygen deficient plane of the CuO chains.\cite{PhysRevLett.60.2198,PhysRevB.39.9667,0953-8984-2-6-021,PhysRevB.43.10422,Nieminen19911577,fermischool1,RevModPhys.66.841,ybcoapl} 
Due to the high positron specificity to oxygen vacancies around the Cu(1) sites the momentum of the annihilating pair measured with (C)DBS is highly sensitive on the oxygen deficiency $\delta$ and in turn to the local transition temperature $T_\mathrm{c}$ in YBa$_{\text{2}}$Cu$_{\text{3}}$O$_{\text{7}-\delta}$.

 \section{Experimental Methods}
 
\subsection{Single-Crystalline YBCO Thin Films}
A single-crystalline YBCO film with a thickness of 230(10)\,nm was grown epitaxially on a routinely cleaned, single crystalline (001) oriented SrTiO$_3$ (STO) substrate (5$\times$5$\,mm^2$) by Pulsed Laser Deposition (PLD).\cite{RevModPhys.72.315,hammerlnature} 
A KrF laser with a fluence of 2\,J/cm$^2$  was used restricting pulse energy to 750\,mJ and pulse frequency to 5\,Hz. 
Deposition took place at around 760\,$^\circ$\,C at a defined oxygen pressure of 0.25\,mbar. 
After deposition, the film was annealed at 400\,$^\circ$\,C in a 400\,mbar O$_{\text{2}}$ atmosphere for oxygen loading. 
Afterwards, a critical temperature of $T_\mathrm{c}=90$\,K was determined by electron transport measurements. 
The single-crystallinity of the grown film was confirmed by X-Ray Diffraction (XRD).
A linear equation reported in ref. \cite{Benzi2004625} was used to determine the overall oxygen deficiency $\delta$ from $\Theta$-$2\Theta$-scans. 
In the as prepared state $\delta=0.191$ and after the heat treatment described below  $\delta=0.619$ were identified.

\subsection{(Coincident) Doppler Broadening Spectroscopy}

In (C)DBS, high-purity Ge detectors are used to measure the Doppler shifted energy $E$ = 511\,keV$\pm\Delta E$ of $\gamma$-quanta emitted from positrons annihilating with electrons.
The Doppler-shift $\Delta E=pc/2$ predominantly results from the (longitudinal) momentum $p$ of the electron  ($c$ is the velocity of light).
The energy of a single annihilation quantum is analyzed in DBS whereas in CDBS\cite{PhysRevLett.77.2097} both annihilation quanta are detected in coincidence using a collinear detector set-up. 
We evaluated the DBS spectra by calculating the line shape parameter $S$, which is defined as the fraction of annihilation quanta with $\Delta E < 0.85$\,keV of the Doppler broadened annihilation line. 
Applying CDBS enhances the peak-to-background ratio and the respective spectra $I(\Delta E)$ were extracted by an algorithm described elsewhere \cite{Pikart201461}.

It was shown that (C)DBS is highly sensitive to the oxygen deficiency $\delta$ in YBa$_{\text{2}}$Cu$_{\text{3}}$O$_{\text{7}-\delta}$:
a higher $\delta$ leads to a less broadened annihilation line and hence to an increase of $S$.\cite{Smedskjaer198856,PhysRevB.36.8854,PhysRevLett.60.2198,1402-4896-1989-T29-019} 
This correlation was found to be linear in our YBCO films.\cite{ybcoapl} 
The present experiments were performed at the CDB-spectrometer\cite{1742-6596-443-1-012071, Gig17} at the high-intensity positron beam NEPOMUC\cite{beam}. 
A variable positron implantation energy $E_+$ in the range between 0.3 and 30\,keV enables in situ depth dependent investigations at temperatures up to 900\,$^\circ$C.

\section{Experiments}
\subsection{In-situ DBS During Tempering}
\label{sec:exp1}
\begin{figure}[t!]
\includegraphics[width=0.495\textwidth]{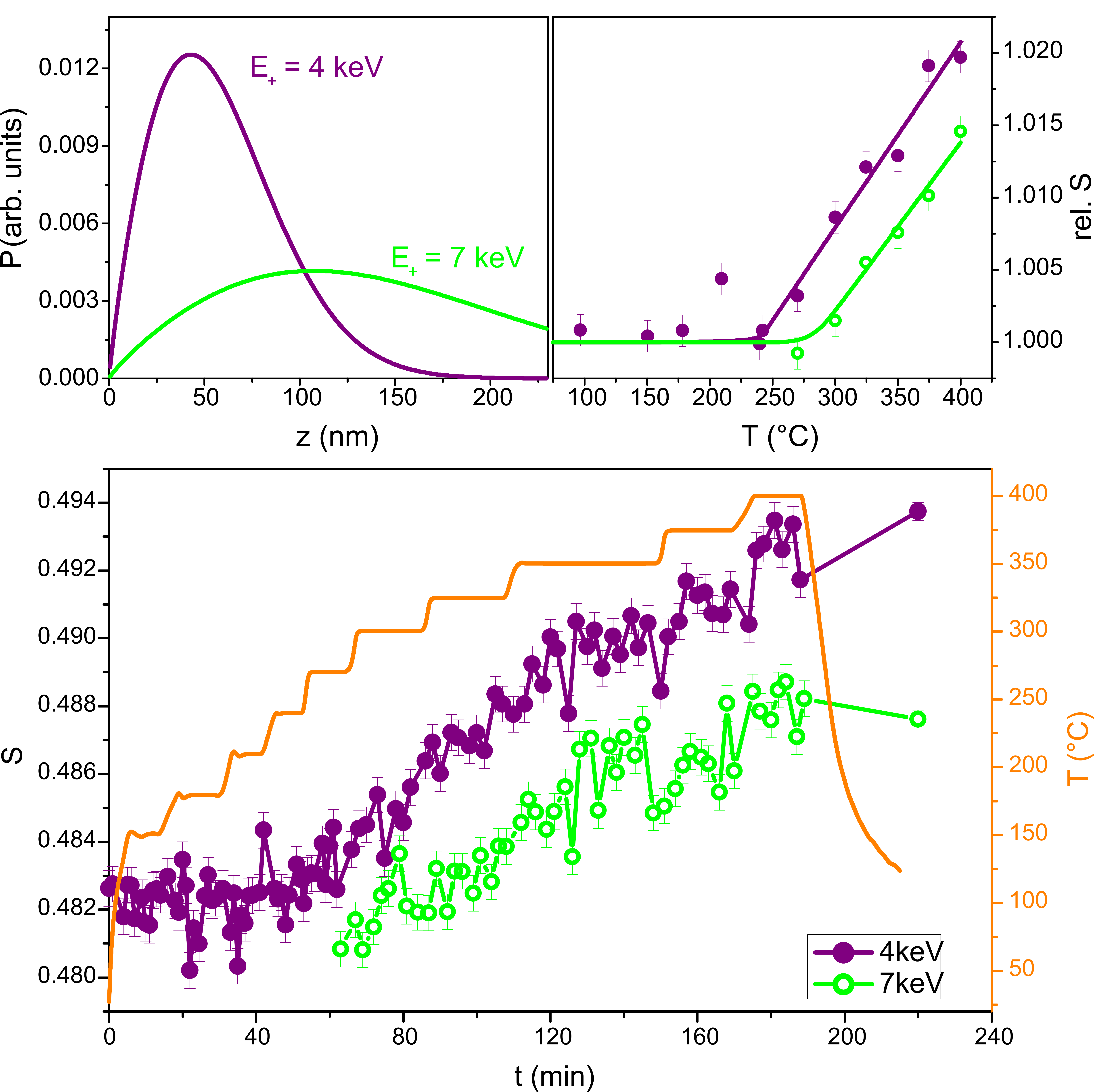}
\caption{In-situ DBS during tempering: a) Positron implantation profiles $P(z,E_+)$ as function of the depth $z$ for the used incident positron energies $E_+$, b) S-parameter and temperature $T$ as function of process time $t$, and c) normalized $S$ as function of temperature $T$ with guides to the eye.}
\label{figure1}
\end{figure}

For studying the oxygen diffusion we performed DBS in situ during elevating the temperature and by alternately switching the positron implantation energy $E_+$ between 4 and 7\,keV. 
The respective Makhovian implantation profiles $P(z,E_+)$, as plotted in fig. \ref{figure1}a),  were calculated with the material dependent parameters obtained from an interpolation over the mass density\cite{param}
and by accounting for the boundary condition of a continuous transmission at the YBCO/STO interface. 
At 4\,keV, positrons exclusively probe the bulk of the YBCO film whereas at 7\,keV the probed region is closer to the interface, and about 9.0\,\% of the positrons actually annihilate in the STO substrate according to the evaluation of the depth dependent measurements discussed in Section\,\ref{sec:exp3}.
As shown in fig.\,\ref{figure1}b) the temperature  $T$ was increased stepwise up to 400\,$^\circ$C within a total measurement time of around 3\,h. 

For both probed depth regions, i.e. at the incident energies $E_+=$\,4 and 7\,keV, the heat treatment led to an increase of the S-parameter (see $S(t)$ in fig.\,\ref{figure1}b)). 
The last S-value reached at each temperature step normalized to the initial S-parameter is plotted as function of $T$ in fig.\,\ref{figure1}c). 
We find at both probed energies a linear $S(T)$ dependence above the respective onset temperatures which is slightly above 240\,$^\circ$C at $E_+ = 4$\,keV and around 280\,$^\circ$C at $E_+ = 7$\,keV. 
As observed in our previous study\cite{ybcoapl} this change of the S-parameter is attributed to the increase of $\delta$.
The S-parameter $S_\mathrm{STO}$ of the STO substrate remains unchanged during tempering as obvious from depth dependent DBS presented in Section\,\ref{sec:exp3}. 
 The mean, i.e. not depth resolved, change of the oxygen deficiency was determined by complementary XRD studies on the film before and after the heat treatment yielding an overall increase of $\delta$ from 0.191 to 0.619. 
However, the quantitative analysis of in-situ DBS at elevated temperatures shows that the variation of $S$ significantly depends on the probed depth region in the YBCO film suggesting a non-constant $\delta$ changing with $z$.
It is noteworthy that $S$(4\,keV) increases by 2.4\,\% whereas $S$(7\,keV) rises only by 1.3\,\% (see fig.\,\ref{figure1}b)).
Since at $E_+=4$\,keV all and at $E_+=7$\,keV a positron fraction of  91.0\,\% annihilates in the YBCO film, an assumed constant $S_\mathrm{YBCO}(z)$ would imply an increase of $S$(7\,keV) by 0.91$\cdot$2.4\,\%=2.2\,\%. 
Hence, a deeper analysis of the non-constant S-parameter in the YBCO film $S_\mathrm{YBCO}(z)$ is expected to provide detailed depth dependent information of the oxygen diffusion process. 

\subsection{CDBS -- Structural Changes}
\label{sec:exp2}
\begin{figure}[t!]
\includegraphics[width=0.485\textwidth]{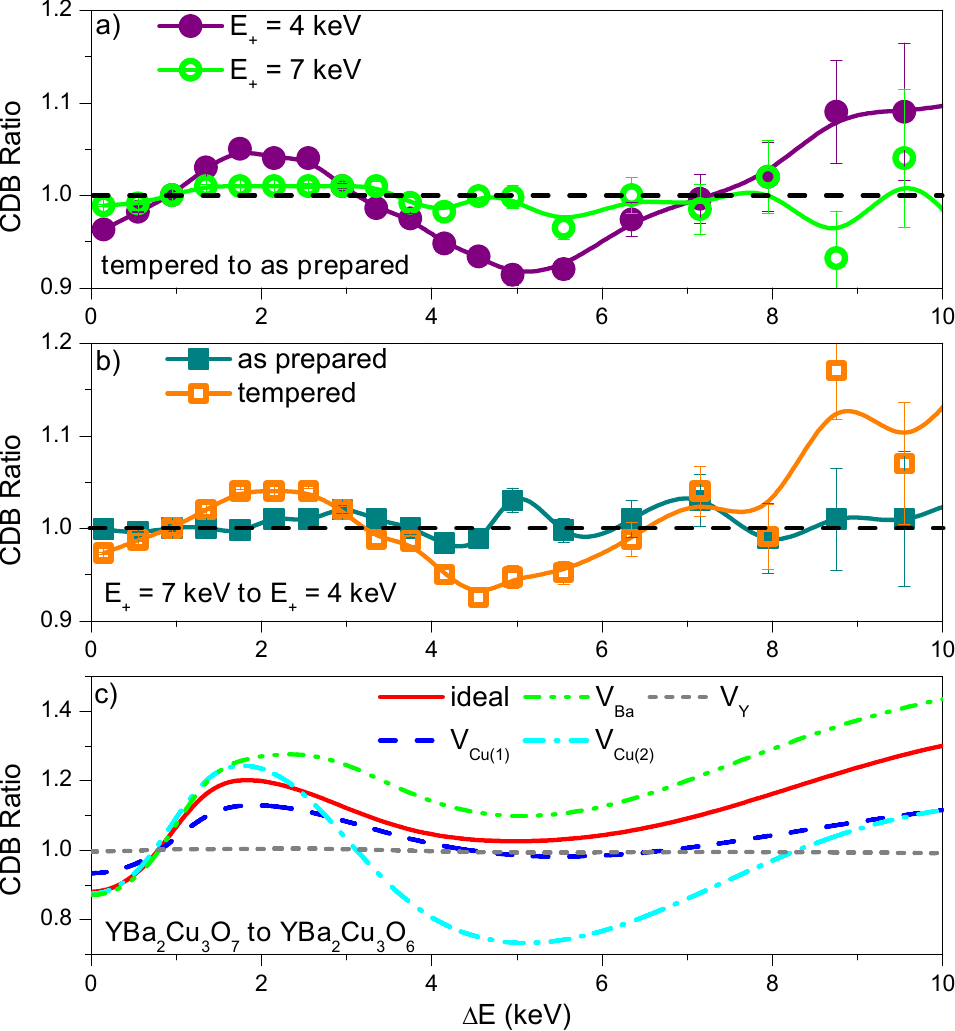}
\caption{CDB ratio curves: a) Tempered to as prepared state for positron implantation energies $E_+=4$\,keV and 7\,keV (see fig.\,\ref{figure1}a) for probed depth region),
b) $E_+=7$\,keV to $E_+=4$\,keV  for as-prepared and tempered  state, and 
c) calculated ratio curves of YBa$_2$Cu$_3$O$_{7}$ to YBa$_2$Cu$_3$O$_{6}$ for various positron states.}
\label{figure2}
\end{figure}

In order to observe the depth dependent change in $\delta$ we recorded  CDB spectra in the as prepared state and after tempering at two different depth regions probed with positron implantation energies of $E_+=4$\,keV and 7\,keV.
 For this purpose we compared the tempered with the as prepared state by evaluating the CDB ratio curves  $R_1(\Delta E) = I_{\mathrm{temp}}(\Delta E, 4$\,keV$)/I_{\mathrm{a.p.}}(\Delta E, 4$\,keV$)$ and $R_2 = I_{\mathrm{temp}}(7$\,keV$)/I_{\mathrm{a.p.}}(7$\,keV$)$ as shown in fig. \ref{figure2}a).
(For reasons of clarity, the argument  $\Delta E$ is omitted in the following). 
Alternatively, the same spectra are analyzed using the ratio curves obtained at different energies for the as prepared and tempered state, respectively, $R_3  = I_{\mathrm{a.p.}}(7$\,keV$)/I_{\mathrm{a.p.}}(4$\,keV$)$ and 
$R_3  = I_{\mathrm{temp}}(7$\,keV$)/I_{\mathrm{temp}}(4$\,keV$)$ (see fig. \ref{figure2}b)). 
In addition, theoretical ratio curves of defect-free YBa$_2$Cu$_3$O$_{7}$ to YBa$_2$Cu$_3$O$_{6}$ as well as those for various metallic vacancies V$_{\mathrm{x}}$ as potential annihilation sites in YBCO, which were calculated for earlier studies\cite{ybcoapl}, are plotted in fig. \ref{figure2}c). 
  
The ratio curve at 4\,keV $R_1$ exhibits a signature characteristic for oxygen-rich crystals as the theoretical ratio curve of YBa$_2$Cu$_3$O$_{7}$ to YBa$_2$Cu$_3$O$_{6}$  in fig.\,\ref{figure2}c) displays. 
In addition, $R_1$ takes values below unity for 3\,keV $< \Delta E <$ 7\,keV, which is attributed to the presence of vacancies V$_{\mathrm{Cu(2)}}$ in the CuO$_2$ planes.
Other remaining positron states, which might contribute to  the signature of $R_1$, cannot be further identified due to the relatively small differences seen between the calculated ratio curves for the various other annihilation sites.
At higher implantation energy $R_2$ only slightly differs from unity which demonstrates that changes are clearly smaller in the depth region probed at $E_+=7$\,keV. 
In the as prepared sample the similarity of $R_3$ with unity shows no evidence for a depth dependent variation of annihilation sites, and positrons annihilating in the STO substrate barely, if at all, affect the CDB signatures. 
After tempering, however, $R_4$ behaves similar to $R_1$ showing a CDB signature characteristic for oxygen-rich crystals and  emerging V$_{\mathrm{Cu(2)}}$ vacancies. 
Hence, we conclude that after tempering $\delta$ decreases towards the interface. 

\subsection{DBS -- Depth Dependent Investigations}
\label{sec:exp3}
\begin{figure}[b!]
\includegraphics[width=0.495\textwidth]{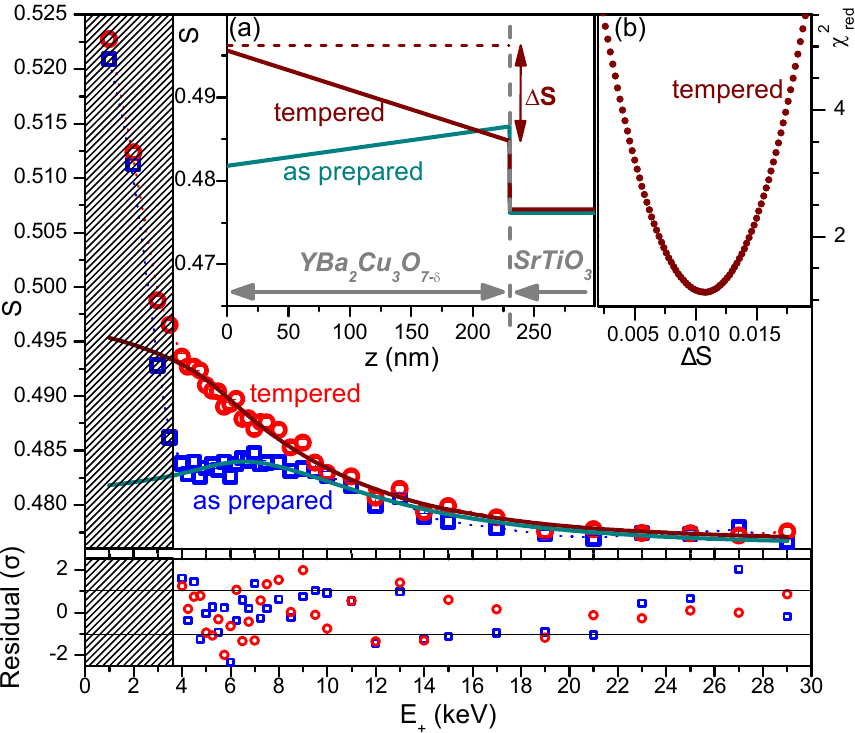}
\caption{S-parameter as function of positron implantation energy $E_+$ before and after tempering. 
The solid lines represent fits yielding $S$(z) as plotted in inset (a) with the respective goodness of the fit $\chi_{red}^{2}(\Delta S)$ with $\Delta S = S(0\mathrm{\,nm})-S(230\mathrm{\,nm})$ in the tempered sample (inset (b)).
For $E_+ < 4$\,keV positron states at the surface affect the data.}
\label{figure3}
\end{figure}

The depth dependent change of the oxygen deficiency is studied in more detail by analyzing the S-parameter as function of positron implantation energy $S(E_+)$ recorded before and after tempering. 
In general, the measured S-parameter as shown in fig.\,\ref{figure3} can be described by a superposition of different positron states with characteristic values at the surface $S_{\mathrm{surf}}$, in the YBCO film $S_\mathrm{YBCO}$ and in the STO substrate $S_\mathrm{STO}$.
For higher implantation energies $E_+$, a significant fraction of positrons annihilates in the substrate with $S_\mathrm{STO}$ whereas for $E_+ < 4$\,keV positrons also annihilate at the surface with $S_{\mathrm{surf}} > 0.52$. 
In the as prepared state, a plateau between 4 and 8\,keV indicates the predominant annihilation in the YBCO film. 
After tempering $S$ increases in this region as expected from the results obtained by the in-situ measurements at higher temperature.
Both, $S_\mathrm{STO}$ and $S_{\mathrm{surf}}$ hardly change during tempering. 
However, detailed information on the depth profile $S_\mathrm{YBCO}(z)$ in the YBCO film can be extracted from  $S(E_+)$. 
 
We performed least-square fits of the $S(E_+)$ curves by considering the depth distribution of annihilating positrons at each energy $E_+$ using the positron implantation profiles $P(z,E_+)$.
Based on previous experimental studies\cite{ybcoapl} positron diffusion plays no significant role along the normal of the YBCO film since positrons either stick at the oxygen deficient plane of the CuO chains or are trapped in metallic vacancies. 
The sharp decrease of $S(E)$ at the surface as observed for $E_+ < 4$\,keV (see fig.\,\ref{figure3}) also indicates a very short positron diffusion length. 
The positron diffusion length $L_{+,\mathrm{STO}}$ in STO was treated as free parameter and was found to be 175\,nm. 
Subsequently, we folded $P(z,E_+)$ with hypothetical $S(z)$ profiles in order to obtain the best agreement with the measured $S(E_+)$ curve. 
It was not possible to get a reasonable fit by assuming $S_\mathrm{YBCO}(z)$ to be constant in the tempered YBCO film. 
Instead, applying a linear dependence of $S$ on $z$ was found to be needed to obtain excellent agreement with the measured data of both, the as prepared and tempered state. 
The fit results of $S(E)$ and $S(z)$ are depicted as solid lines in fig.\,\ref{figure3} by minimizing $\chi_{red}^{2}(\Delta S, S_{\mathrm{STO}},L_{+,\mathrm{STO}})$ as shown in fig.\,\ref{figure3}b. 
In the as prepared state $S_\mathrm{YBCO}(z)$ slightly increases towards the YBCO/STO interface.
After tempering, however, $S_\mathrm{YBCO}(z)$ is on average  higher and decreases towards the interface within the found range $\Delta S$. 
These results allow us to determine $\delta (z)$ profiles for both YBCO films.

\section{Discussion}

The non-constant $S(z)$ found in the YBCO film displays a depth dependent $\delta (z)$. 
We calculated the respective $\delta (z)$ depth profiles using the linear correlation between $S$ and $\delta$ derived in a previous study \cite{ybcoapl}. 
The $S$-$\delta$ calibration was done by extrapolation from the reference values determined by XRD $\delta = 0.191$ and 0.619 for the as prepared and tempered state and the respective values of $S$ obtained by averaging $S(z)$ in the YBCO film  (see fig.\,\ref{figure3}). 

\begin{figure}[b!]
\includegraphics[width=0.494\textwidth]{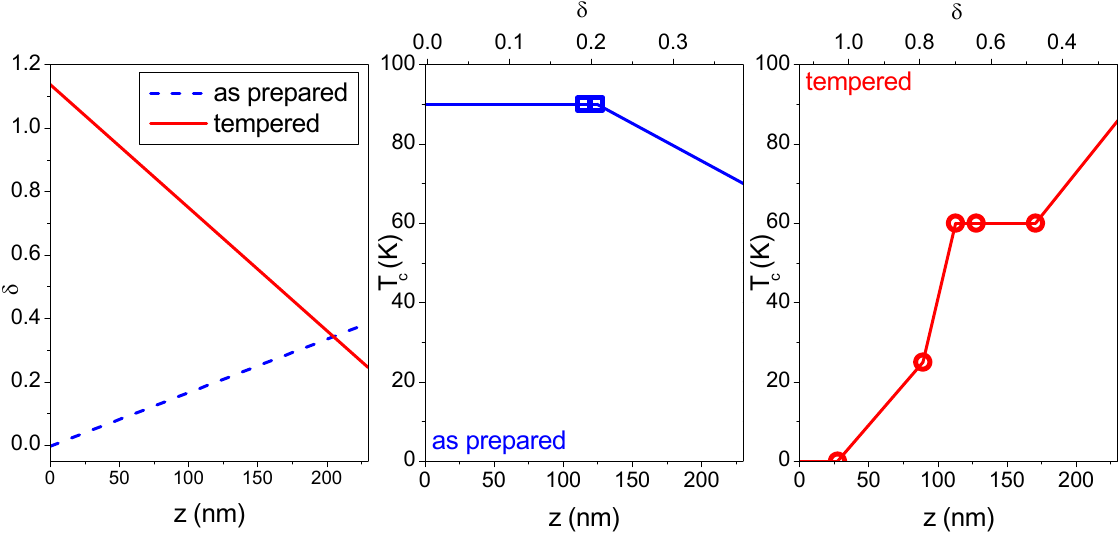}
\caption{Oxygen deficiency $\delta$ and inferred critical temperature $T_\mathrm{c}$ as function of depth $z$ in the as prepared and tempered YBCO film.}
\label{figure4}
\end{figure}

As shown in fig. \ref{figure4} in the as prepared state we observed an increase of $\delta$ from 0.0 at the surface to 0.4 at the interface and after tempering a decrease from 1.1 to 0.2. 
The linear dependence reflects a steady state solution of the diffusion equation for oxygen and hence, displays a state of thermodynamic equilibrium. 
The oxygen deficiency $\delta (0{\mathrm{\,nm}})$ represents the value where oxygen exchange between the film and atmosphere is in equilibrium. 
The as prepared state was achieved in an oxygen atmosphere of 400\,mbar  at 400\,$^\circ$C which leads to oxygen loading of the YBCO film. 
Hence the maximum oxygen content is reached at the surface $\delta (0{\mathrm{\,nm}})\approx 0$ and decreases towards the interface (higher $\delta$).
The heat treatment in vacuum led to oxygen unloading of the tempered sample yielding $\delta (0{\mathrm{\,nm}})\gtrsim1.1$. 
According to the observed $\delta (z)$ behaviour the oxygen concentration increases towards the interface, i.e.\,the oxygen deficiency near the surface is significantly higher than deep in the film close to the interface.

As obvious from the measured $S(T)$ dependence (cf. fig. \ref{figure1}c)) applying a higher temperature leads to an increase of the oxygen loss and hence to an increase of $\delta$.
The step-like behaviour of $S(t)$ (see fig. \ref{figure1}) shows that temperature induced changes end within minutes above 350\,$^\circ$C.
Hence thermodynamic equilibrium is estimated to be reached within the same time scale.  
Our results suggest that in this regime the maximum temperature is decisive for the finally reached mean value of $\delta$. 
The strong increase of $\delta$ towards the surface leads to a lower overall oxygen content inside the YBCO film.
At the interface, however,  $\delta (230{\mathrm{\,nm}})$ slightly decreases by around 0.15 during the heat treatment. 
This (small) effect might indicate that oxygen diffused from the STO substrate into the YBCO film during tempering. 
 
The depth dependence found for $\delta (z)$ implies a depth variation of the critical temperature $T_{\mathrm{c}}$ since it is directly correlated to the oxygen deficiency $\delta$. \cite{PhysRevB.73.180505,PhysRevB.74.014504,Matic2013189} 
The $T_{\mathrm{c}}(z)$ behaviour can hence be derived by interpolating the values determined by measurements on reference samples.
As expected for  the as prepared state, $T_{\mathrm{c}}(z)$ changes only slightly since $\delta$ covers mostly a range where  $T_{\mathrm{c}}(\delta)\approx 90$\,K, and becomes lower near the YBCO/STO interface (see fig.\,\ref{figure4}). 
After tempering, however, $\delta$ varies in a large range that leads to a more complex $T_{\mathrm{c}}(z)$. 
In the near surface region we expect insulating behaviour and for 100\,nm $ < z < $ 200\,nm HTS with $T_{\mathrm{c}}\approx 60$\,K. 
Closer to the interface, $T_{\mathrm{c}}$  increases further. 
Thus in applications of such HTS films the operating temperature can be used as driving factor for the width of the superconducting layer.  
It is noteworthy that the obtained $T_{\mathrm{c}}$-characteristics follow from the measured $\delta (z)$ depth profile only, whereas in transport experiments just a mean value of $T_{\mathrm{c}}$ is accessible neglecting lateral inhomogeneities in the YBCO film and in which the contacting of the samples might effect the measurement\cite{ybcoapl}.

The present results demonstrate that the mobility of oxygen atoms in YBCO plays an important role for the preparation of high-quality films using PLD. 
In several studies, typical diffusion lengths for oxygen in YBCO bulk samples were determined elsewhere. 
A heat treatment at 430\,$^\circ$C for 20 minutes yielded diffusion lengths of 10\,nm along the $c$-axis and 6\,$\mu$m along the $a$- and $b$-directions.\cite{PhysRevB.51.8498} 
However, in the thin film samples of the present study the oxygen diffusion along $c$ is expected to be faster by roughly one order of magnitude since thermodynamic equilibrium was already reached after several minutes of tempering above 350\,$^\circ$C. 
An extraordinarily high oxygen mobility along the $c$-axis has been reported in other studies on YBCO films produced by magnetron sputtering\cite{xixx89} and laser ablation\cite{PhysRevB.51.8498}. 
Moreover, we observed an onset temperature at around 240\,$^\circ$C for oxygen diffusion, which is lower compared to values published for bulk YBCO, e.\,g., 350\,$^\circ$C observed by electric resistance measurements\cite{PhysRevB.47.3380} or 400\,$^\circ$C by positron annihilation spectroscopy\cite{PhysRevB.43.10399}. 
In other studies on YBCO films similarly low onset temperatures around 250\,$^\circ$C have been determined by a combination of oxygen tracer diffusion and secondary ion mass spectroscopy.\cite{PhysRevB.51.8498} 
Our CDBS results suggest that the observed high mobility of oxygen atoms and the low onset temperature for oxygen diffusion might be connected to V$_{\mathrm{Cu(2)}}$ vacancies in the CuO$_2$ planes.

Finally, we discuss the homogeneity of $\delta$ in thin single-crystalline YBCO films with a thickness of several 100\,nm as characteristic for laser-ablated samples. 
Along the $c$-axis we have observed a thermal equilibrium state with a linear $\delta (z)$ depth profile determined by the surface value $\delta (0{\mathrm{\,nm}})$ and the oxygen content at the YBCO/STO interface $\delta (230{\mathrm{\,nm}})$. 
The value of $\delta (0{\mathrm{\,nm}})$ can be adjusted by partial oxygen pressure in the atmosphere and by temperature. \cite{PhysRevB.43.10399}
Assuming that the STO substrate provides an ideal homogeneous reservoir of oxygen, only the maximum temperature should affect the exchange rate of oxygen at the YBCO/STO interface and hence the value of $\delta (230\mathrm{\,nm})$.
However, more detailed insights into this process and  the properties of the interface is beyond the scope of the present study. 
In practically identical thin film YBCO samples we found lateral inhomogeneities of $\delta$ along the direction of $a$ and $b$.\cite{ybcoapl} 
According to the kinematics discussed above, we estimate that a tempering time in the order of 200\,hrs is required to reach a laterally homogenous distribution of oxygen. 
Since local structural variations of the interface between STO substrate and the YBCO would affect the oxygen exchange future studies with the scope on the surface homogeneity of the STO substrate could provide further useful information. 
Possibly, the homoepitaxial deposition of a single-crystalline STO layer by PLD prior to the growth of the YBCO film with adjustable defect densities in the intermediate STO layer \cite{PhysRevB.79.014102,PhysRevB.81.064102,PhysRevLett.105.226102} might improve the homogeneity of the interface. 
Incorporating an oxygen diffusion barrier at the interface would lead to a constant equilibrium depth profile $\delta (z) = \delta (0{\mathrm{\,nm}})$. 
Therefore, such a layer is expected to simplify significantly the preparation of YBCO films with a homogeneous and precisely defined oxygen deficiency $\delta$.\\

\section{Conclusion and Outlook}
In this study we investigated the diffusion related properties of the oxygen deficiency $\delta$ in epitaxial single-crystalline YBa$_{\text{2}}$Cu$_{\text{3}}$O$_{\text{7}-\delta}$ thin films.
The averaged $\delta$ could be determined by XRD, and the mean  $T_{\mathrm{c}}$ was obtained from transport measurements.
By applying (C)DBS with a variable energy positron beam, we succeeded in revealing the depth distribution of the oxygen content which in turn unveils the depth dependent critical temperature $T_{\mathrm{c}}$. 
Thus in future applications it has to be considered that changes in the ambient temperature affect the width  and hence the current density of the superconducting layer.
In situ measurements at elevated temperature allowed us to gain insights into the kinematics of oxygen atoms. 
An onset temperature for oxygen diffusion was found slightly above 240\,$^\circ$C reflecting the high mobility of oxygen along the $c$-axis.
The depth distribution of oxygen in the thermodynamic equilibrium after preparation and after the heat treatment  were shown to be driven by the oxygen exchange at the surface and to lesser extent at the YBCO/STO interface.
The oxygen content near the surface can be manipulated relatively easy by the ambient pressure and temperature, whereas the control of the interface processes is more demanding. 
Solving this issue is key for further improving the quality of single-crystalline YBCO films in terms of a precisely defined homogeneous oxygen deficiency. 
The availability of such samples exhibiting unprecedented quality is estimated to be highly relevant for a better understanding of fundamental phenomena and for technical applications of the HTS YBCO.

\section*{Acknowledgments}
Financial support from the DFG within project  TRR\,80 and from the  BMBF projects nos. 05K13WO1 and 05K16WO7 is gratefully acknowledged. The authors thank M. Leitner for helpful discussions.

\end{document}